\documentclass{elsart}
\usepackage{graphicx}
\usepackage{bm} 
\usepackage{amssymb,amsmath,color,cite,multicol}
\journal{Nuclear Physics A}

\begin{document}
\begin{frontmatter}

\title{Binding vs. scattering length: $\eta$ in light nuclei}
\author[a]{J.A. Niskanen}
\author[b]{H. Machner}

\address[a]{Department of Physics, PO Box 64, FIN-00014 University of
Helsinki, Finland }
\address[b]{ Fachbereich Physik,
Universit\"{a}t Duisburg-Essen, Lotharstr. 1, 47057 Duisburg, Germany}

\date{\today}

\begin{abstract}
The  possibility of etamesic
nuclei remains an open problem in nuclear physics until now.
Various calculations give contradictory predictions even
for the lightest real nucleus $^3$He. In this paper we present
the connection of the binding energy and width to the
complex scattering length for
$s$-states in heavier nuclei than this
in the hope that, with knowledge
of the final state interaction this could be useful in searches
of possible bound states. It is seen that for a consistent
analysis also the effective range should be considered.
\end{abstract}
\end{frontmatter}

\section{Introduction}
In many unstable hadronic systems perhaps the only way to
get information on their structure and interactions is the
final state interaction in their formation process and
associated decays, enhancement or decrease {\it vs.}
free undistorted final state. Production slightly above
the free threshold can yield also information on possible
bound states below the threshold, especially if they are
close to the threshold, {\it i.e.} weakly bound. This can
be seen in the energy dependence \cite{Watson, Migdal} and
described by the final state low-energy scattering parameters.

However, the cross section alone cannot distinguish whether
the interaction can or cannot support a bound state.
A textbook example is singlet $S$-wave $NN$ scattering.
It was necessary to indulge the difference in coherent
neutron scattering off para- and ortohydrogen molecules
to extract the sign of the large scattering length, which
in turn showed that the interaction is not binding\cite{bw}.
In most systems, in particular in case of spin-0 particles,
this kind of extra information is not available. However,
as final state interaction analyses anyway give {\it some}
information, it is conceivable that this would be useful
in experimental searches for bound or quasibound states.
In the latter case the problem can be far from trivial,
since the state, even if "bound", can be wide and
correspondingly the the low-energy scattering parameters
would be complex.

An example of recent interest is the possibility of
$\eta$-nuclear bound states. Numerous calculations
exist, which disagree with each other completely especially
for the lightest "real" nuclear systems with $^3$He and
$^4$He \cite{Wilkin, Wycech, Rakityansky, Fix1, Haider1, Liu,
Chiang, Haider2}. Some of them indicate binding whereas most
don't, while a general consensus is that by carbon binding
should exist. Ref. \cite{Sibirtsev} presents an overview
of the confusion and a new fit for $(|a_{\rm R}|,a_{\rm I})$
summarizing the experimental efforts.

Only a few experiments
have been performed. One class of experiments produces the
$\eta$-meson at rest in a quasi-free transfer reaction.
In a second step the $\eta$ interacts with a nucleon thus
forming a resonance $N^*(1535)$ which can decay back to its
entrance channel or, with 50$\%$ probability, into a nucleon
and a pion. Since the $\eta$ is at rest, these two final state
particles are emitted almost back to back. The experiment by
the GEM collaboration \cite{Budzanowski09} claimed a 5$\sigma$
effect in studying the $p+^{27}$Al$\to ^3$He$+\pi^-+p+$X
reaction at a beam momentum for which the intermediate $X=\eta$+(A-2)
is almost at rest. In an experiment employing photoproduction
the existence of $\eta$-mesic ${\rm{^3He}}$ was claimed to have
been observed in the reaction $\gamma + ^3$He$\to \pi^0+p+X$ using
the photon beam at MAMI  \cite{Pfeiffer04}.  Similarly as in
the previous experiment a two step process was assumed but only
the pion was measured and not the other nucleons. It has, however,
been pointed out \cite{Krusche10} on the basis of new high
statistics data for the excitation function of the reaction
$\gamma+^3$He$\to \pi^0+p$+X that the data of Ref. \cite{Pfeiffer04}
do not permit an unambiguous determination of the existence of a
$\eta{\rm{^3He}}$-bound state, because nucleon resonances
produce opening angle dependent structures in excitation functions
and subtraction of excitation functions for different opening
angles can produce artificial structures almost anywhere.

Inclusive experiments searching for $\eta$-mesic
nuclei at BNL \cite{Chrien88} and LAMPF \cite{Lieb88}
by using a missing-mass technique in the ($\pi^+, p$)
reaction reached negative or inconclusive results.
Later it became clear that the peaks are not necessarily
narrow and that a better strategy of searching for
$\eta$-nuclei is required as for instance applied in
Ref. \cite{Budzanowski09}. Furthermore, the BNL experiment
was in a region far from
the recoilless kinematics, so the cross section is
substantially reduced \cite{Hirenzaki07}.

Another class of experiments searched for $\eta$-mesic nuclei
in final state interaction. Intensive studies were dedicated
esp. to the $p+d\rightarrow \eta+^3$He reaction
\cite{Mayer,Smyrski07,Mersmann07}. The $\eta^4$He final state
has been studied in $d+d$ interactions making use of unpolarized
beams \cite{Wronska05, Frascaria94, Willis97} as well as
polarized beams \cite{Budzanowski09b}. The very large momentum
transfer tends to make direct production of $\eta$ mesons
more difficult with larger nuclei.
The heaviest system studied so far for final state interactions
is  $\eta^7$Be produced in $p+^6$Li reactions
\cite{Scomparin93, Budzanowski10}. In this case
 there are only two data points
at about 13 and 19 MeV above the threshold, so that no attempt
for a final state interaction is possible, yet.

With reasonable assumptions of the Watson-Migdal theory
\cite{Watson,Migdal} final state studies can give
estimates for the imaginary value of the scattering length
and the absolute value of its real part \cite{Sibirtsev}.
However, the sign of the latter would be crucial as a
tell-tale of a bound state.
Still, even $|a_{\rm R}|$ could give indications of the
value of the binding energy, {\it provided it exists},
useful for experiments searching for such states.
Further useful information would be expectations of the
width of such states. The aim of the present paper is to
continue to heavier light nuclei the investigation of
Ref. \cite{Sibirtsev2} for $^3$He on the relation between
binding and the low-energy scattering parameters.

The paper presents the minute amount of formalism next
and then the results for representative mass distributions
of three light nuclei.

\section{Formalism}
There is not much actual new formalism in this paper. Rather
the aim is a numerical extension of Ref. \cite{Sibirtsev2}
to heavier nuclei than $^3$He exposing more some details.
The basic idea is
to start from a simple optical model with a potential
proportional to the density profile of the nucleus, use it
to calculate the complex binding energy and scattering
parameters separately and combine them to a common contour
plot in the $(a_{\rm R},a_{\rm I})$ plane. This presentation
of the binding energy and width as a function of the
complex scattering length is not necessarily trivial.
However, due to the shape independence of nuclear forces
with this phenomenology a connection between the basically
distinct observables can be considered as better justified
than their direct connection to the potential, independent
on the validity of the simplest
impulse approximation optical model.

For specificity (and to facilitate a comparison to the
impulse approximation), the potential can be expressed as
\begin{equation}
V_{\rm opt} = -4\pi (V_{\rm R} + i V_{\rm I}) \rho(r)/
(2\mu_{\eta N})\, ,
\end{equation}
with $\mu_{\eta N}$ the reduced mass of the $\eta N$ system.
Here the nuclear density $\rho$ can be varied from nucleus
to nucleus and for each nucleus the strength parameters are
freely varied to get a sufficient coverage of  the
$(a_{\rm R},a_{\rm I})$ plane. It should be stressed that
we are not predicting any absolute strength of the potential
as in the model works referred above. The main thing is
the numerical connection of binding energies and widths to
the scattering parameters, so that if the latter can be
extracted from data, then a preliminary estimate could be
obtained for the former. Although both are in some ranges
sensitive to the potential parameters, in the spirit of the
shape independence of $NN$ forces, one might expect a
relatively density profile independent connection. Indeed,
in Ref. \cite{Sibirtsev2} it was checked that the relation
between $(V_{\rm R},V_{\rm I})$ and  $(a_{\rm R},a_{\rm I})$
was robust against changes in the density profile.
In contrast to the free variation, within the optical model
(as {\it e.g.} in Ref. \cite{Wilkin}) the strength would be
related to the elementary $\eta N$ scattering length
as $V_{\rm R(I)} = A a_{\eta N,{\rm R(I)}} \
$ with $A$ the
atomic number of the nucleus.

The scattering program is fairly standard even with a complex
potential. This involves solving the Schr\"odinger equation
with the proper asymptotic boundary condition
\begin{equation}
R_l(r)  \sim j_l(kr + \delta_l)
\end{equation}
with $k = \sqrt{2\mu_{\eta A} E/\hbar^2}$ and
$\mu_{\eta A}$ is the reduced mass of the $\eta$-nuclear
system.
The binding solutions are obtained searching by
iteration for poles in the homogeneous
Lippmann-Schwinger integral equation (in
configuration space)
\begin{equation}
R_l(r) = -ik \frac{2\mu_{\eta A}}{\hbar^2}
\int_0^\infty j_l(kr_< )\, h_l^{(1)}(kr_> ) V(r')
R_l(r')r'^2 dr'\, ,
\end{equation}
equivalent to the Schr\"odinger equation. The Green
function arguments are $r_<\; (r_> )$ the smaller (larger)
of $r$ and $r'$. For $s$-wave bound states this reduces to
\begin{equation}
R_l(r) = -\kappa \frac{2\mu_{\eta A}}{\hbar^2}
\int_0^\infty \frac{\sinh (\kappa r_< )}{\kappa r_<}
\, \frac{\exp(\kappa r_> )}{\kappa r_>}\, V(r')\,
R_l(r')\, r'^2\, dr'\, ,
\end{equation}
where now $\kappa = \sqrt{-2\mu_{\eta A} E/\hbar^2}$.

The convergence was
good except for real potentials with very small binding energy
($\leq 0.1$ MeV) where the wave functions are much more extensive
than the potential range. This case could reasonably be
considered as essentially the zero binding limit with also
extremely large cross section.
Convergence stopped also in case of very
large widths $(\Gamma/2 \geq 250)$ MeV, in which case the wave
functions tend to be of shorter range than the potential.
The latter case is
certainly not of experimental interest (with binding still
at most in low tens of MeV or rather a few MeV).

The $s$-wave scattering parameters
are defined as is standard for mesons by
\begin{equation}
q \cot \delta = \frac{1}{a} + \frac{1}{2} r_0 q^2
\label{definition}
\end{equation}
so that for a real attractive potential
$a_{\rm R} < 0$ means binding (we shall later bring up
a more exact condition). Experiments extract so far only the
scattering length $a$, but it is notable that the effective
range $r_0$ is of the same order in the range of most interest.
Therefore, its experimental determination (or inclusion
of theoretical predictions by hand) in analyses would also be
of interest and importance.

\section{Results}
As a representative example the most detailed discussion is
given to $^{12}$C where binding is unanimously assumed. For
this the modified harmonic oscillator of Ref. \cite{atomic} is
used as the density profile
\begin{equation}
\rho(r) = 0.17\, [1 + 1.15\,( \frac{r}{1.672} )^2 ]
\exp{[-(r/1.672)^2]}\; {\rm fm}^{-3}
\end{equation}
with the normalization to the atomic number as
$4\pi\,\int_0^\infty \rho\, r^2\, dr = 12$
and $r$ given in fm.

\begin{figure}[tb]
\centering
\includegraphics[width=0.8\textwidth]{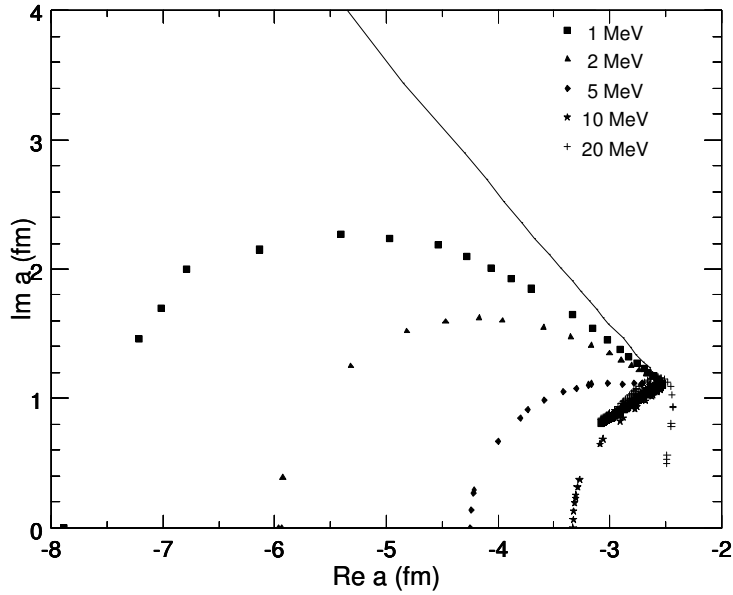}
\caption{The $s$-wave binding energy
$E_{\rm B} = - E_{\rm R}$ contours for 1, 2, 5, 10 and
20 MeV in the complex $(a_{\rm R},a_{\rm I})$ plane.
The line shows the zero energy, {\it i.e.} above it there
is no binding as explained in the text.}
\label{Ereal}
\end{figure}

\begin{figure}[tb]
\centering
\includegraphics[width=0.8\textwidth]{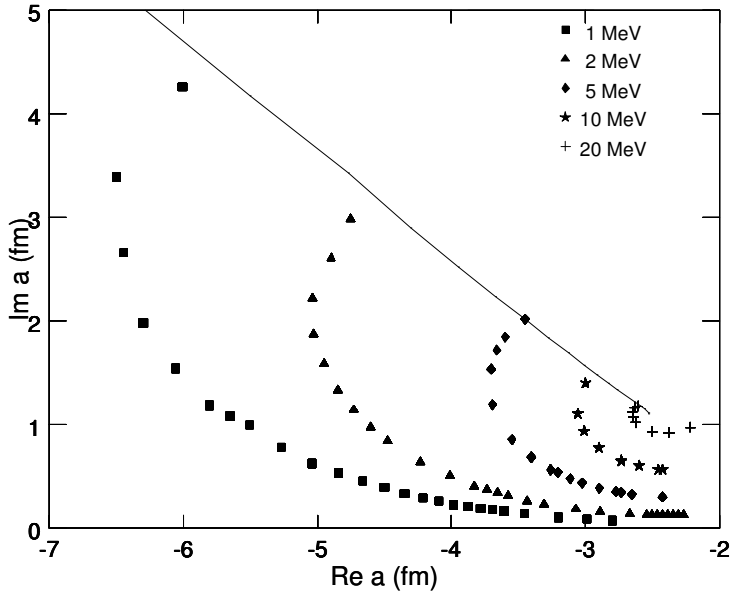}
\caption{The same as Fig. \ref{Ereal} but for the imaginary part
 of the bound state energies $-E_{\rm I}$, {\it i.e.} half widths.
}
\label{Eimag}
\end{figure}

The basic results are given in Figs. \ref{Ereal} and
\ref{Eimag}, where the binding energies (defined as
$E_{\rm B} = - E_{\rm R} > 0$)
and half-widths $ \Gamma/2 = - E_{\rm I}$
are presented as contour points for 1, 2, 5, 10 and 20 MeV.
The basic criterion for a printed point is that the
deviation from the value is less than 0.05 MeV, though also
a linear interpolation or extrapolation has been used in some
more sensitive instances.

At least for real potentials and small binding
in Fig. \ref{Ereal} the results follow well the trend
$a \sim E_{\rm B}^{-1/2}$ dictated by general arguments
\cite{general,baru}. In fact, starting from the defining
equation (\ref{definition}) one can derive a relation between
the binding energy and low-energy parameters \cite{joachain}.
This can be generalized to the complex case as
\begin{equation}
1/a = -\sqrt{-2\mu_{\eta A}E/\hbar^2}
    -  r_0\, \mu_{\eta A}\, E/\hbar^2
\label{relation}
\end{equation}
with $\mu_{\eta A}$ the reduced mass of the system. This relation
was found to be amazingly accurate predicting the value of $a$
well for binding energies up to $|E|\approx 10$ MeV and even
beyond (a few percent for the real part and about ten percent
for the imaginary; $E$ and $r_0$ taken from the
calculation). This success can be attributed to the inclusion
of a non-zero effective range $r_0$ and can be considered as
another indication of its importance.

\begin{figure}[tb]
\centering
\includegraphics[width=0.8\textwidth]{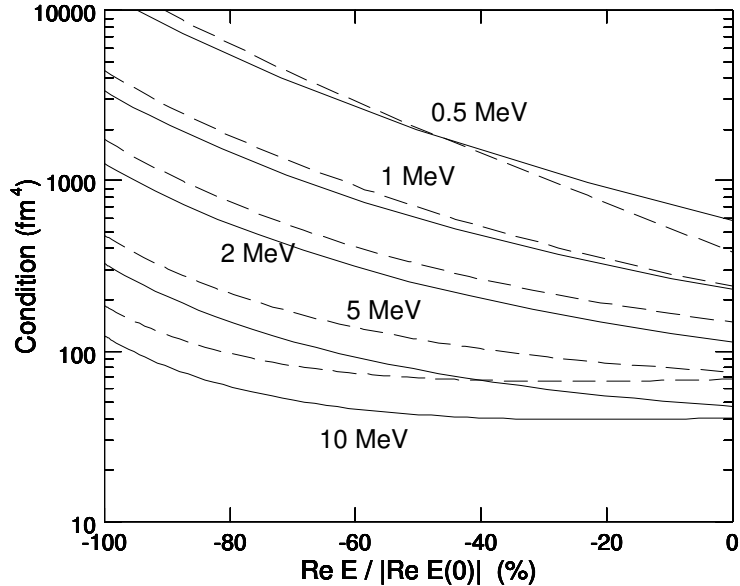}
\caption{The condition (\ref{condition}) (solid)
{\it vs.} its first term (dashed) for five real potentials
giving maximum binding energies indicated for real potentials.
The horizontal axis is the percentage fraction
$E_{\rm R}(V_{\rm I})$ of  $E_{\rm R}(0)$.
}
\label{figurecondition}
\end{figure}

Furthermore, it may be noted that for numerically bound
points the condition \cite{Sibirtsev}
\begin{equation}
 \mathcal{R} [a^3(a^\ast - r_0^\ast)] > 0
 \label{condition}
\end{equation}
was well satisfied, while the simpler rule
$|a_{\rm R}| > a_{\rm I}$
without the effective range, given {\it e.g.} in
\cite{Haider2},
extended for (wide) virtual states relatively far above
threshold, {\it i.e.} the inequality was satisfied also
for unbound states. This latter condition is equivalent
to keeping only the first term in expression (\ref{condition}),
so a comparison of the conditions is numerically possible.
As seen in Fig. \ref{figurecondition}, also
expression (\ref{condition}) could remain positive
beyond the bound region, although it is mostly smaller,
and both conditions
decrease by an order of magnitude for decreasing $E_{\rm R}$
for each given $V_{\rm R}$. Therefore, both conditions
turn out to be necessary but not sufficient, though
(\ref{condition}) is more precise. In fact, it also
extends less to the unbound region.

With complex potentials the dependencies become nontrivial.
As befits strong interactions, the strength of the potential
$V$ (real or imaginary) does not go to $E$ or $a$ linearly.
However, in the former case the dependence of both $E_{\rm R}$
and $E_{\rm I}$ is anyway monotonous once the corresponding
other part is kept constant. In contrast, for a given
$V_{\rm R}$ $a_{\rm I}$ is {\it not} monotonous with respect
to $V_{\rm I}$. This behaviour results in the two "branches"\
of the $E_{\rm R}$ plots. The upper branch (starting from left
for $E_{\rm R} < 10$ MeV and potential real) could be
considered as a "weak potential" part.\footnote{It
should be noted that this "branching" does not
refer to the nomenclature in analyticity properties. Rather
the question is about backbending of the curves. In the
present results there is still a unique correspondence
between the {\it complex} scattering lengths and energies.}
A general
rule of thumb is that the imaginary potential, absorption,
behaves like repulsion, although the above mentioned
non-monotonousness means that, in a way as inelasticity,
it eats its own effect
out at some stage. While $E_{\rm I}$ grows with
increasing $V_{\rm I}$, the real part $E_{\rm R}$
decreases, eventually to no binding. This also means that
for a given constant binding $E_{\rm R}$ a stronger
$V_{\rm R}$ is needed, when $V_{\rm I}$ increases.
In Fig. \ref{Ereal} the lower "strong potential"
part is particularly dictated by the repulsive effect of the
imaginary part of the potential. There an even mesh of
potential strengths gives an increasingly dense accumulation of
points.
In this region a small change of the
complex $a$ (in particular $a_{\rm I}$) can produce sizable
changes in the binding and width.
However, this region of $|a| \approx$ 2--3 fm is also of most
interest concerning both theoretical predictions and experiments
at least in the helium case \cite{Sibirtsev}.

\begin{figure}[tb]
\centering
\includegraphics[width=0.8\textwidth]{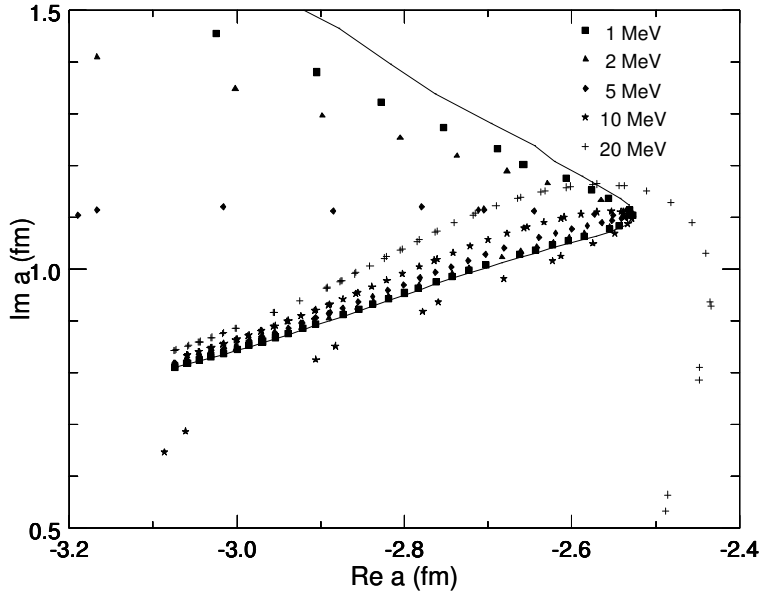}
\caption{Magnified detail of Fig. \ref{Ereal}
}
\label{mini}
\end{figure}

Therefore, Fig. \ref{mini} shows a magnified view of the
backbending region. It can be seen that with
increasing binding energy the upper "weak potential" branch
comes down, whereas the lower "strong potential" part slowly
bulges up. Therefore the opening angle between the weak
and strong potential branches gets smaller until at about
9 MeV binding energy they switch over. Consequently,
the zero binding curve (solid) in the lower branch is not
actually a limit of possible bound states (as the
upper branch of the solid line is). The "weak" coupling
states get below it. It is also noteworthy that
the strong coupling results as curves are not very far from
each other (again in contrast to the weak coupling) and
consequently not far from the lower zero energy line.
In fact the 1 and 2 MeV results are indistinguishable in
that case. Further it may be noted that the most "eastern"
point of the zero binding curve is $(-2.525 + i\, 1.102)$ fm.

\begin{figure}[tb]
\centering
\includegraphics[width=0.8\textwidth]{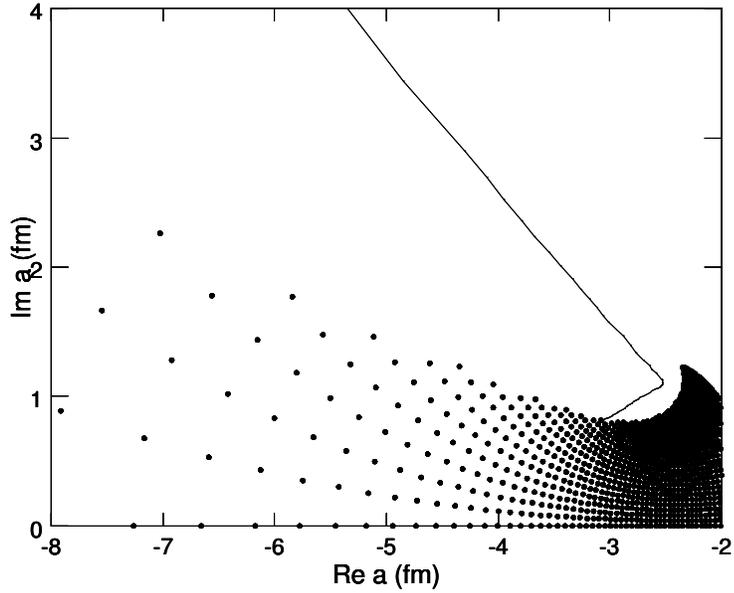}
\caption{The region where $|E_{\rm R}| > |E_{\rm I}|$.
The solid curve is the zero binding limit.
}
\label{bigger}
\end{figure}

In addition to the absolute values of binding energies and
widths, of paramount experimental interest is their relative
magnitude. Typically for experimental recognition of a
bound state one would hope the width or half-width
to be less than the binding energy for distinguishing a state
from continuum. For this purpose Fig. \ref{bigger} shows by
black points the region for which $|E_{\rm R}| > |E_{\rm I}|$.
This belongs to the realm of "weak" coupling results.
Quite clearly the real part of the scattering length in general
should be larger than the imaginary part.

\begin{figure}[tb]
\centering
\includegraphics[width=0.8\textwidth]{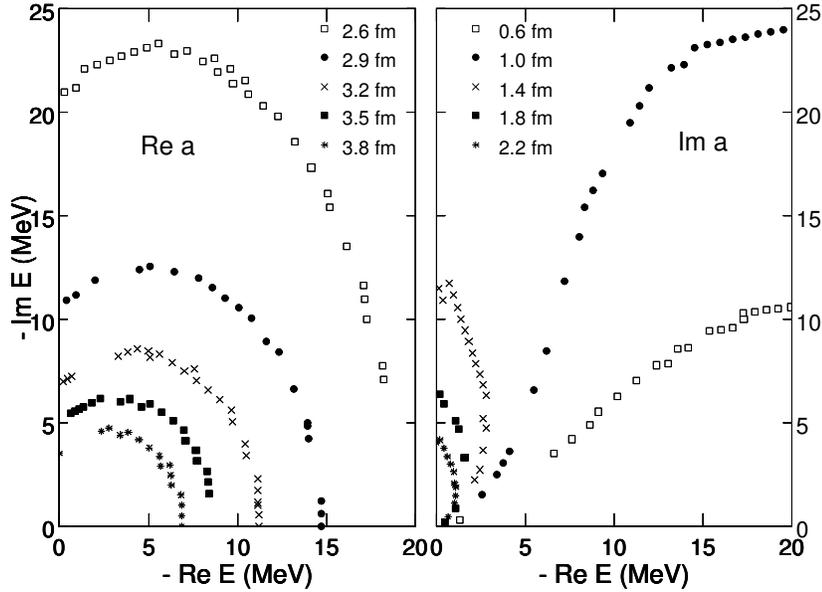}
\caption{The real and imaginary parts of the scattering length
presented by contours in terms of the binding energy
$E_{\rm B} = - E_{\rm R}$ and half-width
$\Gamma /2 = - E_{\rm I}$.
}
\label{avse}
\end{figure}

Sometimes it might be desired that possibly available
binding energies would be used to predict
or calculate as a test the low energy scattering parameters.
Since without a too elaborate analysis or a detailed
scattering calculation the effective range is not
then known, a unique determination of the scattering
length is not possible by Eq. (\ref{relation}) alone.
Therefore, in Fig. \ref{avse} we give the reverted result
for carbon, namely the components of the complex $a$ as
contours in the plane defined by the binding energy and
half-width. Here the real part of the scattering length
has double-valued relation to the width. Also $a_{\rm I}$
curves would return  to the right side of the $E_{\rm I}$
axis for very large values of $E_{\rm I}$, {\it i.e.}
for width values above 100 MeV, clearly of no physical
interest.

\begin{figure}[tb]
\centering
\includegraphics[width=0.8\textwidth]{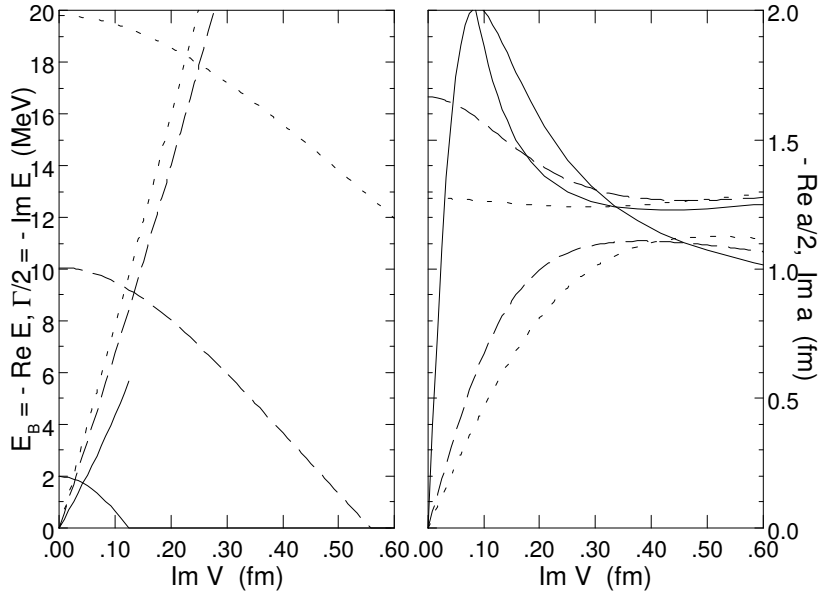}
\caption{Dependence of the $s$-state
binding energies and half-widths
(left panel) and the complex scattering lengths (right panel)
on $V_{\rm I}$. Three real potential strengths  $V_{\rm R}$
are used, corresponding to binding energies of 2 (solid),
10 (dashed) and 20 MeV (dotted) in the real case,
respectively, as explained in the text. The imaginary parts
are curves starting from the origin, while the real parts
start from finite values.
}
\label{vieffect}
\end{figure}

As the connection between the complex $V$ and $a$ clearly
is not trivial, therefore, before going on to other
nuclei it may be of interest to study their
interdependence in some more detail, in particular the
somewhat unintuitive effect of $V_{\rm I}$.
Along with the effect on the binding energy,
this is done in Fig. \ref{vieffect}, where these observables
are shown as functions of $V_{\rm I}$ for three
different values of $V_{\rm R}$,
which would give binding energies of 2 (solid),
10 (dashed) and 20 (dotted) MeV for the real case. These
are obtained with $V_{\rm R}$ equal to 0.22, 0.37 and 0.50 fm,
respectively, though, as stressed before, not too much
weight should be associated with the absolute strengths
of the phenomenological potential. However, these are
provided here for reproducibility of the present results.

The
binding energies $-E_{\rm R}$ decrease fast with increasing
$V_{\rm I}$ as expected (obviously starting from 2, 10 and
20 MeV for $V_{\rm I}=0$). The dependence of $E_{\rm I}$
on $V_{\rm I}$ is locally very well linear and steep.
In the behaviour of the scattering
lengths the first characteristic feature is the saturation
of $a_{\rm I}$, even turning slowly downwards, as mentioned
earlier for the rising absorption potential strength.
(These curves are recognized as starting from the origin.)
Further, the corresponding curves for $- a_{\rm R} /2$ are
shown. Also here the relative constancy of $a_{\rm R}$
is remarkable for the strongly imaginary potentials.
The curves of the real part of
the scattering length for all values of
$V_{\rm R}=0$ appear to converge towards about -2.6 fm,
apparently a sort of a "soft black sphere" limit.
(It may be noted that the rms radius for this mass
distribution is 2.44 fm.) These latter
features cause the backbending of the constant energy
curves and also the strong accumulation of points in Fig.
\ref{Ereal}. After all, it must be difficult to present
$E$ graphically as a function of nearly constant
components of $a$. It is also worth noting that in
the region of $V_{\rm I}$ where $a_{\rm I}$
reaches about 75 -- 80\% of its maximum and above,
$E_{\rm I}$ (half-width) exceeds $E_{\rm R}$.

\begin{figure}[tb]
\centering\includegraphics[width=0.8\textwidth]{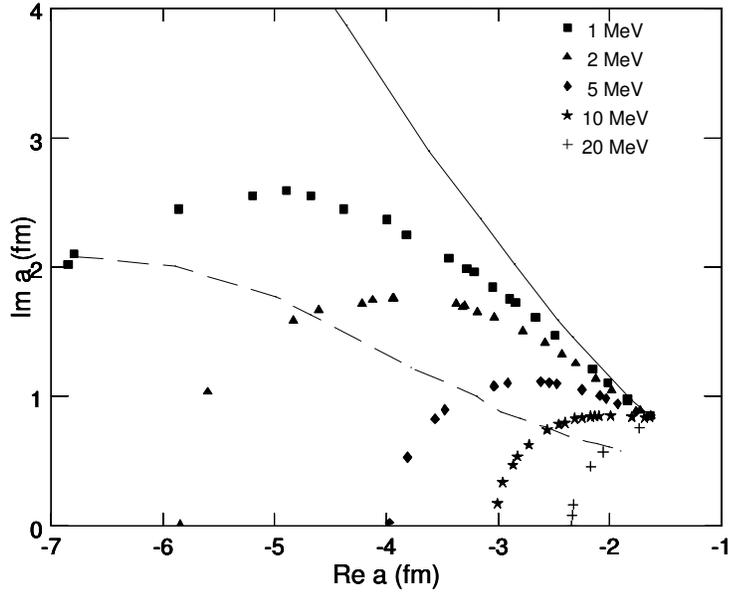}
\caption{The same as Fig. \ref{Ereal} but for the nucleus $^4$He.
Below the dashed line  $|E_{\rm R}| > |E_{\rm I}|$
}
\label{Ehe4}
\end{figure}

Carbon in most models would support binding and it is also
of great interest to get predictions for
more controversial lighter nuclei
with possible final state interaction fits. Therefore we
used the three parameter Fermi distribution
\cite{atomic}
\begin{equation}
\rho (r) = 0.24\, \frac{1+0.517 r^2/0.964^2}{1 + \exp((r-0.964)/0.322)}\; {\rm fm}^{-3}
\end{equation}
for the density profile of $^4$He to get similar estimates
in a much lighter and more controversial case. $^3$He was studied
in Ref. \cite{Sibirtsev2} and the question raised again in
Ref. \cite{anke}. Considering that even the nuclei are rather
different, as seen in Fig. \ref{Ehe4}
the results are surprisingly similar to those of carbon.
In practice only the turning point for weak and strong
potentials has changed from about $(-2.5 + i\, 1.1)$ fm to
$(-1.5 + i\, 0.9)$ fm. For a real potential the scattering length
corresponding to 1 MeV binding changes from $-7.9$ fm to $-7.4$ fm.
These, of course, reflect primarily the difference in effective
ranges, which for carbon varies roughly between 2.5 fm ("weak"
potential) and 1.5 fm ("strong") and for helium 1.7 fm and 1 fm,
respectively, and is, of course, complex. With eq. (\ref{relation})
the differences would, thus, be as expected.

\begin{figure}[tb]
\centering
\includegraphics[width=0.8\textwidth]{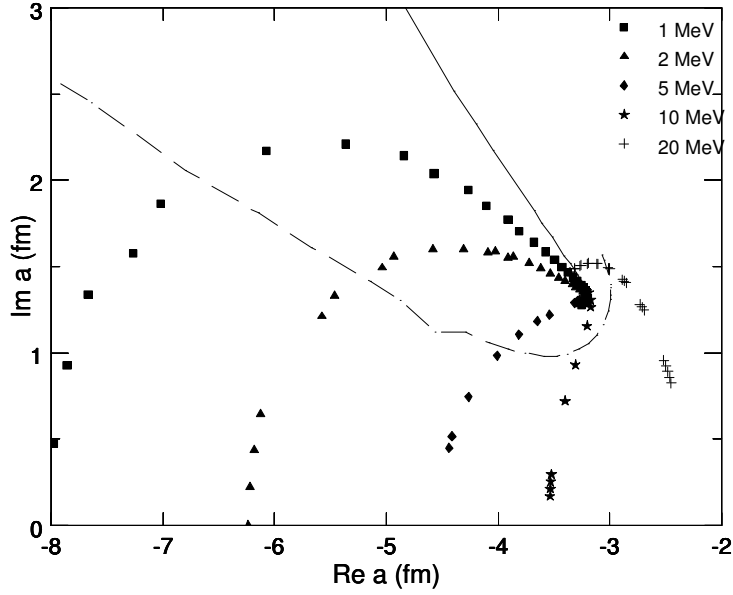}
\caption{The same as Fig. \ref{Ereal} but for the nucleus
$^{24}$Mg. Below the dashed line  $|E_{\rm R}| > |E_{\rm I}|$
}
\label{EBMg24}
\end{figure}

As an extension to heavier nuclei we choose to use also
a three parameter Fermi distribution with $A=24$ ($^{24}$Mg
from Ref. (\cite{atomic}))
\begin{equation}
\rho (r) = 0.17\, \frac{1-0.163 r^2/3.108^2}{1 + \exp((r-3.108)/0.607)}\; {\rm fm}^{-3} \, .
\end{equation}
This should be close enough a possible observation of
a bound $\eta$ state in $^{25}$Mg of Ref.
(\cite{Budzanowski09}). The results for the binding energy
contours are shown in Fig. \ref{EBMg24}. Especially up to
the binding of 10 MeV they are very similar to those
of carbon. Only this time the turning point has shifted
to the left by about 0.6 fm. Taking the position of the
minor peak of Ref. (\cite{Budzanowski09}) below the $\eta ^{25}$Mg
threshold for it face value as 13 MeV binding and
estimating the half width from the data distribution to
be 5 MeV would then correspond to the complex scattering
length of $a \approx -3.1 + i\, 0.6$ fm (with also
$r_0 \approx 1.6 - i\, 0.6$ fm; the calculated width and
effective range not shown by figures). Now, if only one
had data on final state interactions in agreement with
these, the additional data would corroborate the
interpretation as a bound state.

From the figures \ref{Ereal}, \ref{Ehe4} and  \ref{EBMg24}
one can see that for more extensive distributions of larger
nuclei (increasing effective range) the energy contours
shift to the left towards larger values of $|a|$. The
same, of course, holds for inclusion or exclusion of a
positive effective range in analyses. This effect has
actually been seen in analyses of Refs.
\cite{Smyrski07,Mersmann07} for the $\eta^3$He final state.
Smyrski {\it et al.} get $a = [\pm(2.9\pm 2.7)+i(3.2\pm 1.8)]$
fm (without $r_0$), whereas Mersmann {\it et al.} quote
a dramatically different value
$a = [\pm(10.7\pm 0.8_{-0.5}^{+0.1})+i(1.5\pm 2.6_{-0.9}^{+1.0})]$
fm (with $r_0$). One should
note, however, that there is also a difference in taking
into account the resolution smearing. Similarly
Ref. \cite{Budzanowski09b} gives as the best fit
$a = [\pm(3.1\pm 0.5)+i(0\pm 0.5)]$ fm without $r_0$
and $a = [\pm(6.2\pm 1.9)+i(001\pm 6.5)]$ fm with
the effective range term included in the low energy
expansion. Of course, these different results arise
from the same physics, the same bound or unbound state.
One may look at this effect also in another way.
For the same values of $a_{\rm R}$ the binding energies
should get bigger with a finite positive $r_0$.
So (assuming a negative $a_{\rm R}$, {\it i.e.} a bound state)
the first value of Ref. \cite{Budzanowski09b} would
give a binding $E_{\rm B} = 10\pm 3$ MeV from Fig.
\ref{Ehe4} instead of 4 MeV without the second term
in (\ref{relation}). However, the second value of $a$ may
be more consistent with reality, though one cannot say
that from the fit itself. That would give a binding
of about two MeV from Fig. \ref{Ehe4}.

So it seems likely that the effective range
can be essential in a consistent analysis. In particular,
a real first principles calculation for finite sized
nuclei and the relation (\ref{relation}) cannot be
accommodated without a finite $r_0$ and the same argument
can be raised for data analyses. A single number for
$r_0$ is not enough, since it is a
function of the complex strength.
Therefore, to facilitate its (at least approximate)
use we give its real and imaginary components
fitted as a second order bipolynomial form of the complex
scattering length as
\begin{equation}
{\rm Re}\, r_0 = c + d\, a_{\rm R} + e\, a_{\rm I}
+ f\, a_{\rm R}^2 + g\, a_{\rm I}^2 +h\, a_{\rm R}\, a_{\rm I}
\label{fitform}
\end{equation}
and correspondingly for ${\rm Im} \, r_0$,
with the coefficients $c - h$ given in Table \ref{efrfit}
for each nucleus and component. For numerical and physical
reasons the fit was constrained to the regions $|a| < 8$ fm
as in the figures, excluding the very large scattering
lengths which would be inproportionately weighted by this
form and for energies for which
$|E_{\rm I}| < |E_{\rm R}$ (the
region of physical interest with distinguishable peaks).
From the figures this would mean a binding of at least
1 MeV in the real case.
Smaller binding might still be well
described, since the second term in (\ref{relation})
should then become small.
Discarding the excluded region improved
the $\chi^2$ significantly.

Thus all three observable quantities are interrelated.
Once two are
known, by shape independence quite well established in
this work, the remaining one is determined without
knowledge of actual potential details, such as its
absolute strength. The moderate dependence on just the
density profile may be a reasonable assumption as
a starting point and could be taken from {\it e.g.}
the present results.

\begin{center}
\begin{table}[tb]
\centering \caption{Coefficients of the powers or products
of the expansion (\ref{fitform}) for the complex effective
range ($a$ and $r_0$ in fm) for the three nuclei considered.
 }
\vspace{3mm}
\label{efrfit}
\begin{tabular}{l|c|c|c|c|c|c}
Quantity  & Constant & $a_{\rm R}$ & $a_{\rm I}$  & $a_{\rm R}^2$ &
$a_{\rm I}^2$ & $a_{\rm R}a_{\rm I}$  \\
\hline
Re $r_0(^{4}$He) &   0.54215 &  -0.28931 &  0.11454 &  -0.015985 &
 0.0041414 &   0.018879 \\
Im $r_0(^{4}$He) & 0 & 0 &  -0.36374 & 0 &  -0.011609 &  -0.048794 \\
\hline
Re $r_0(^{12}$C) &  -1.1515 &  -1.1416 & 0.30317  &  -0.090601 &
  0.21724 &  0.10477 \\
Im $r_0(^{12}$C) & 0 & 0 &  -1.207 & 0 &  -0.22788 &  -0.23799 \\
\hline
Re $r_0(^{24}$Mg) &   -2.282  &  -1.5557 & 0.33591 &  -0.11709 &
  0.26911 &  0.11353 \\
Im $r_0(^{24}$Mg) & 0 & 0 &   -1.5419 & 0 & -0.092067 & -0.25479 \\
\hline
\end{tabular}
\end{table}
\end{center}

\section{Conclusion}
In this work a phenomenological connection between the
low energy scattering length and the complex binding
energy of possible eta-nuclear bound states is studied
in a simple but probably realistic model. The purpose
of the
work is that the results may be of use in searches
of these bound states, if more easily accessible
final state data are available to make predictions
where to look for the states. The binding energies are
explicitly presented as contours in the complex $a$
plane for the nuclei $^{12}$C, $^4$He and $^{24}$Mg.
The well established and checked shape  independence
gives smooth systematics from which it is easy to
interpolate and even extrapolate to other nuclei.

The calculations suggest that
for even relatively moderate values of the imaginary
potential and of the imaginary parts of the scattering lengths,
the states can be wide especially compared with the
real depths of the states. In view of also many other
theoretical results,
starting from the elementary $\eta N$ scattering and predicting
negative real parts for the scattering length but with rather
large imaginary parts, the observation of such bound states
might be difficult or even impossible. However, in the
minireview \cite{Sibirtsev} of the
situation a reanalysis of the existing data on $\eta^3$He
final states makes very small values of the imaginary part
plausible, so that also the possible bound states may not
necessarily be as wide as most theoretical works would
indicate.

In our work for $a_{\rm I}$ less than 2 fm with $a_{\rm R}$
larger than, say, 5 fm a bound state should be recognizable.
In the case of more likely smaller scattering lengths
$a_{\rm I} < 1$ fm would be necessary.  In this respect the
result $a_{\rm R} = 6.2 \pm 1.9$ fm
and $a_{\rm I} = 0.001 \pm 6.5$ fm of Ref. \cite{Budzanowski09b}
is quite interesting and suggestive.
 The relation between $a$ and $E$
(as discussed above and evidenced by Figs. \ref{Ereal},
\ref{Ehe4} and \ref{EBMg24})
is very robust against potential differences
even between different nuclides over a wide range.
 Therefore, due to this shape
independence one may trust the results to be valid
by interpolation also
for the experimentally interesting $A = 7$ nuclei.

\section*{Acknowledgements}\label{sec:Acknowl}

We thank Ch. Hanhart for useful discussions.
This work was partly supported by the DAAD and Academy of Finland
exchange programme projects 50740781 (Germany) and 139512 (Finland).

\end{document}